\documentstyle[12pt]{article}

\newcommand{\be}{\begin{equation}}
\newcommand{\ee}{\end{equation}}
\newcommand{\er}{\end{eqnarray}}
\newcommand{\ea}{\end{eqnarray}}
\newcommand{\br}{\begin{eqnarray}}
\newcommand{\ba}{\begin{eqnarray}}

\begin{document}

\thispagestyle{empty} 
$\phantom{x}$\vskip 0.618cm

\begin{center}
{\Huge {\bf Soldering and Mass Generation in Four Dimensions}} 
\vspace{1cm}
$\phantom{X}$\\
{\Large Rabin Banerjee${}^{b}$ and Clovis Wotzasek${}^{a,b}$}\\
[3ex]{\em ${}^{a}$Instituto de F\'\i sica\\Universidade
Federal do Rio de Janeiro\\21945, Rio de Janeiro, Brazil\\}

{\em ${}^{b}$S. N. Bose National Centre for Basic Sciences,\\
Block JD, Sector III, Salt Lake, Calcutta 700091, India.\\}

\end{center}

\vspace{1cm}

\begin{abstract}
\noindent 
We propose bosonised expressions for the chiral Schwinger models
in four dimensions. Then, in complete analogy with the two 
dimensional case, we show the soldering of
two bosonised chiral Schwinger models with opposite chiralities to yield the 
bosonised Schwinger
model in four dimensions. The implications of the Schwinger model or its 
chiral version, as known for two dimensions, thereby get extended to four
dimensions.
\end{abstract}

\newpage

The Schwinger \cite{S} mechanism of generating a photon mass in two space time
dimensions is crucial for our understanding of several aspects of
different models in such dimensions. 
The particular features of the Schwinger model to produce a massive vectorial
mode, while at same time avoiding the Goldstone bosons were, at some point, atributed to be a
consequence of two-dimensional kinematics.  As so these results could not be 
immediately generalized
to higher dimensions. However, by following the soldering mechanism proposed by Stone
\cite{MS}, the massive mode in the Schwinger model was shown by us\cite{BW1}
to be an interference effect between the massles modes of the chiral Schwinger
models \cite{JR}. Since the soldering formalism was very general and did not depend on 
any specific properties of two dimensions, it was felt that the analysis should 
be extended to higher dimensions. This would also provide a scenario for mass generation
in these dimensions. In order to proceed with the analysis it is therefore
essential to have some bosonised expression characterising the Schwinger
model in four dimensions. This was achieved by Aurilia, Takahashi and Townsend\cite{ATT}.
The model had 
important phenomenological consequences.
In particular, it was found to
simulate the effective theory representing
$QCD$ in the large $N$ limit. 
Just as the bosonised Schwinger model in two dimensions was thought to
be the effective theory for $QED_2$, it was argued that the appropriate
generalisation of $QED_2$ to four dimensions done in \cite {ATT}
was a prototype of $QCD$
in the large N limit \cite{Rev}.
More recently,
applications in other contexts like the study of skyrmions
or Josephson arrays have also been mentioned \cite{DT}.  Now the 
Schwinger model in four dimensions was constructed by a direct
lift of the well known bosonised version of the model in two dimensions.
A natural question that arises in this context and which lacked an answer would be the construction 
of the chiral Schwinger model in four dimensions. We could then
interpret this theory to characterise chiral $QCD$ in the large N
limit. Of course we would like
to have the result in the bosonised form since it is in this form the Schwinger
model was written in four dimensions. It may be recalled that the chiral
Schwinger model in two dimensions is iteslf an interesting model \cite{JR}
providing an alternative mechanism for mass generation, which is basically
through the existence of an anomaly. One might be tempted to repeat the
same logic of directly lifting the bosonised version of the chiral Schwinger
model in two dimensions, which is familiar, to four dimensions. Such an
approach, alas, fails. The reason is that in two dimensions the vector and
axial currents are related by an identity so that effectively there is no
difference between them. This is not so in four dimensions. Consequently,
a simple lift of the chiral version of the Schwinger model from two to
four dimensions is not possible.

Recently we \cite{{BW1}, {BW2}} have developed a method by which it is possible to combine
two distinct models into an efective model. This is called the soldering
technique. The idea is to exploit the dual aspects of the symmetry of the
constituent models to construct a new model which effectively hides the
symmetries. In the example at hand it is the chiral symmetry. Thus we expect
that by soldering two chiral Schwinger models of opposite chiralities, the
normal Schwinger model is reproduced. This was shown by us earlier in the
usual case of two dimensions\cite{BW1}. By extending those notions we will construct
here the explicit bosonised forms of the chiral Schwinger models in four dimensions.
However this construction is fundamentally different from the two dimensional
case. It necessitates the introduction of an internal space quite similar to 
what is done for describing electric-magnetic duality symmetric lagrangians \cite{{Z}, 
{BW2}} and must be defined in the Fourier space.
By soldering these chiral models we correctly reproduce the Schwinger model
lagrangian in four dimensions as given in \cite{ATT}.
Comparing with the two dimensional analysis we conclude that exactly the
same mechanism for mass generation occurs also in four dimensions.

To illustrate the concepts in a simple setting, as well as for 
reasons of comparison, the soldering
of two chiral models to yield a vector model in two dimensions is shown.
The first order lagrangians for the gauged chiral bosons are given by \cite{H},

\begin{equation}
{\cal L}_R=-\dot \phi \phi '-\phi '^2 -2e\phi '
(A_0+A_1)-{{e^2}\over 2}(A_0+A_1)^2 +\frac{a\,e^2}{2}A_\mu A^\mu,
\label{eqch1}
\end{equation}
\begin{equation}
{\cal L}_L=\dot \rho \rho '-\rho '^2 +2e\rho '
(A_0-A_1)-{{e^2}\over 2}(A_0-A_1)^2 +\frac{b\,e^2}{2}A_\mu A^\mu,
\label{eqch2}
\end{equation}
These lagrangians correspond to the coupling with the right (R) handed
and left (L) handed pieces, respectively. They were obtained by imposing
the chirality constraints on the standard vector lagrangian so that one
chirality is killed while the other is retained. The parameters $a$ and $b$
reflect the bosonisation or regularisation ambiguity of the chiral determinants.
Imposing Bose symmetry \cite{NB}, which means that both determinants are accounted
by the same sort of ambiguity, we can effectively set $a=b$.
Each of these lagrangians
therefore corresponds to half a degree of freedom. Furthermore, taking the opposite
chiralities into account we can count $\phi$ to have half a degree of freedom
while $\rho$ would carry minus half a degree of freedom.

Next, consider the variation of the lagrangians under the following 
transformations,

\begin{equation}
\delta \phi=\delta \rho=\alpha;~~\delta A_\mu=0.
\label{eqst1}
\end{equation}
We find,

\begin{equation}
\delta {\cal L}_R=2J_R\alpha ',~~~\delta {\cal L}_L=2J_L\alpha ',
\label{eqxyz}
\end{equation}
where the currents are given by,
\begin{equation}
J_R=-(\dot\phi +\phi '+e(A_0+A_1)),~~J_L=(\dot\rho -\rho '+e(A_0-A_1)).
\label{eqxx}
\end{equation}
Introducing the soldering field $B$, transforming as,
\begin{equation}
\delta B=-2\alpha '
\label{eqzz}
\end{equation}
it is found that the soldered Lagrangian,
\begin{equation}
{\cal L}= {\cal L}_R\oplus{\cal L}_L =   {\cal L}_R+{\cal L}_L+B(J_R+J_L)-{1\over 2}B^2,
\label{eqp}
\end{equation}
remains invariant under the combined transformations (\ref{eqst1}) and (\ref{eqzz}). 
Eliminating $B$ in favour of the other variables yields the final effective lagrangian,

\begin{equation}
{\cal L}=\frac 12 \partial_\mu\theta\partial^\mu\theta+ 2\, e\,\epsilon_{\mu\nu}
A^\mu\partial^\nu\theta +(a-1)e^2 A_\mu A^\mu\;\;\; ;\;\;\;\epsilon_{01}=1
\label{schwinger}\end{equation}
where the new field $\theta$ is defined as,

\begin{equation}
\theta=\phi-\rho
\label{field}
\end{equation}
The familiar gauge invariant vector lagrangian is obtained by setting
$a=1$. Inclusion of
the Maxwell term right from the beginning corresponds to the soldering of two 
chiral Schwinger models to yield the normal Schwinger model. All results, of course,
are to be understood in the bosonised version. It might be observed that the
$a=b=1$ parametrisation corresponds to the massless modes in the chiral Schwinger
models\cite{JR}. Thus the massive mode of the Schwinger model is a consequence of the interference of
two massless modes in the chiral models.

Regarding the degree of freedom count we find the soldering of two half
degree of freedom to yield a single degree of freeeom. This is also 
contained in the algebraic relation (\ref{field}) which, for the degree
of freedom count, can be written as $1=\frac 12 - (-)\frac 12$.

We shall now discuss the construction of the bosonised version of the chiral
Schwinger models in four dimensions. The difficulties of a direct lift from its
two dimensional counterpart have been already mentioned. One might think of an
alternative possibility; namely of applying the chirality constraint on the
four dimensional Schwinger model. This also fails since the
peculiar form that the chiral constraints assume in the two dimensional case does not admit
an unambiguous local dimensional extension.
It is our main goal in this paper to propose such a dimensional lift of the chiral Schwinger
model and to show that the interference effects provided by the soldering mechanism
are able to produce the gauge invariant massive mode of the model in \cite{ATT}.

In order to motivate the analysis in four dimensions the results in
two dimensions are
reconsidered by omitting the gauge couplings. This implies the
soldering of two free chiral bosons having opposite chiralities
to yield the free scalar theory. The equation of motion for this
Klein-Gordon field may be expressed in a factorised form,
\be
\triangle \theta=(\partial_0 +\partial_1)(\partial_0 -\partial_1)\theta=0
\label{KG}\ee
where $\triangle$ is the usual Klein Gordon operator.
The factored forms are recognised to represent the equations of motion 
for the chiral bosons, following from (\ref{eqch1}) and (\ref{eqch2}),
\be
\dot\theta\pm\theta'=0
\label{chiral1}\ee
Indeed if we naively multiply the expressions appearing in the
l.h.s. of the above equation we just recover the standard free
scalar lagrangian. The main thrust would therefore be to find
the four dimensional lagrangians leading to a set of equations mimicing
(\ref{chiral1}). With these observations the ensuing analysis 
in four dimensions becomes more transparent.

It is now useful to go over to the $k$-space. To this end let us introduce
an internal
two-dimensional space spanned by the basis\cite{BCW},
$\left\{\hat e_a({\bf k},{\bf x}), \; a=1,2\right\}$,
with $({\bf k},{\bf x})$ being conjugate variables and the orthonormalization condition given as,

\begin{equation}
\label{ortho}
\int d^3{\bf x} \;\hat e_a({\bf k},{\bf x})\hat e_b({\bf k}',
{\bf x})=\delta_{ab}\delta({\bf k}-{\bf k'})
\end{equation}
We choose the vectors in the basis to be eigenvectors of the Laplacian,
$\nabla^2=\partial_m \partial_m$,

\begin{equation}
\label{nabla}
\nabla^2 \hat e_a({\bf k},{\bf x})= -\,\omega^2({\bf k})    \hat e_a({\bf k},{\bf x})
\end{equation}
and the action of $\partial_m$ over the $\hat e_a({\bf k},\bf x)$ basis to be

\begin{equation}
\label{dg230}
\partial_m \hat e_a({\bf k},{\bf x})= k_m \epsilon_{ab}\hat e_b({\bf k},
{\bf x})\;\; ;\;\; \epsilon_{12}=1,
\end{equation}
so that the dispersion relation $k_\mu \, k^\mu = \omega^2 - k_m^2 =0$ holds.
Finally, we define the basic chiral fields carrying the opposite $(\pm)$
chiralities in the momentum space to be the expansion coefficients of
a chiral scalar field in the function space as,

\begin{equation}
\phi^{(\pm)}({\bf x}, t) = \int d^3 {\bf k}  \,\varphi_a^{(\pm)}(t, {\bf k})\,e_a({\bf k},{\bf x}).
\end{equation}
Notice that the internal index disappears upon Fourier transforming
the fields back in the coordinate space.

The expected equations of motion displaying the chiral properties would
therefore be given by,

\begin{equation}
\dot\varphi_a(k)\mp\omega\epsilon_{ab}\varphi_b(k)= 0
\label{motion}\end{equation}
This is the exact analogue of the equation of motion for the free chiral bosons
in two dimensions given in (\ref{chiral1}). Proceeding as before if we multiply
the factors occuring in the l.h.s. of the above equation, we obtain,
after making a Fourier transform to the coordinate space,

\be
\int d^4k(\dot\varphi_a+\omega\epsilon_{ab}\varphi_b)
(\dot\varphi_a-\omega\epsilon_{ac}\varphi_c)
=\int d^4x(\partial_\mu\phi\partial^\mu\phi)
\label{KG1}\ee
which is just the action for the free Klein Gordon field in four dimensions.

We next construct a first order action that would yield the equation of motion (\ref{motion}).
This is given by,

\begin{equation}
\label{chiral_ATT1}
{\cal S}_{\pm}^{(0)}= \int d^4k\left[\pm\omega({\bf k}) 
\dot\varphi_a^{(\pm)}\epsilon_{ab}\varphi_b^{(\pm)}
\, -\, \omega^2({\bf k}) \varphi_a^{(\pm)}\varphi_a^{(\pm)}
\, \right].
\end{equation}
The basic brackets among the fields are easily read off from the
symplectic structure,

\begin{equation}
\{\varphi_a(k), \varphi_b(k')\}=\pm\frac{ 1}{2\,\omega({\bf k})}\epsilon_{ab}\,\delta(k-k')
\label{algebra}\end{equation}
for either chirality. Since this is a first order action the hamiltonian is also
simply read off,

\be
H=\int d^3{\bf k} \,\omega^2({\bf k}) \varphi_a^{(\pm)}\varphi_a^{(\pm)}\label{ham}\ee
From (\ref{algebra}) and (\ref{ham}) the desired result (\ref{motion})
follows. It is in fact possible to explicitly carry out the soldering of
these four dimensional chiral boson lagrangians to get the usual 
Klein Gordon lagrangian. We bypass this to discuss the interacting theory.
Introduce the chiral combinations of gauge fields
${\cal A}^a_{\pm}$ as follows,

\begin{equation}
{\cal A}^a_{\pm}= -\epsilon^{ab}\left(\omega\,
\mbox{}^*A_0^b\pm k_m \mbox{}^*A^b_m\right)
\label{A_field}
\end{equation}
where $\mbox{}^*A^b_\mu$ is the Hodge-dual of a
pair of three form gauge 
fields $A_{\mu\nu\lambda}^a$, defined as,

\begin{equation}
\mbox{}^*A^\mu_a = \frac 12 \epsilon^{\mu\nu\lambda\rho}  A^a_{\nu\lambda\rho}
\;\;\; ;\;\;\; \epsilon^{0123}=1
\label{dual}
\end{equation}

We 
now propose the four dimensional lift of the bosonised version
of the interacting theory by a simple extension of the free
theory (\ref{chiral_ATT1}),

\begin{equation}
\label{chiral_ATT}
{\cal S}_{\pm}= \int d^4k\left[\pm\omega({\bf k}) 
\dot\varphi_a^{(\pm)}\epsilon^{ab}\varphi_b^{(\pm)}
\, -\, \omega^2({\bf k}) \varphi_a^{(\pm)}\varphi_a^{(\pm)}
\, -\, 2g {\cal A}^a_{\pm}\varphi_a^{(\pm)}-\frac{ g^2}{2\,
\omega^2({\bf k})}{\cal A}^a_{\pm}\right]+\frac {a\, g^2}2{\cal S}_{reg},
\end{equation}
where ${\cal S}_{reg}$ manifests the bosonisation ambiguity, quite similarly to
the two dimensional case, and is given by,

\ba
{\cal S}_{reg} &=&  \int d^4k \,\left[ \left(\mbox{}^*A^0_a\right)^2 -
\frac 1{\omega^2}\left(k_m \mbox{}^*A^m_a\right)^2\right]\nonumber\\
&=& \int d^4x\, \left[\mbox{}^*A^m\left(\delta_{mn} -
\frac{\partial_m\partial_n}{\nabla^2}\right)\mbox{}^*A^n +
\mbox{}^*A^\mu\,\mbox{}^*A_\mu\right]
\ea
Notice that if we dimensionally reduce this expression to D=2 the
first piece in the second line vanishes and we get back
the Jackiw-Rajaraman regularisation ambiguity term. Indeed by including the Maxwell term
in (\ref{chiral_ATT}) the corresponding actions represent the bosonised versions of the
chiral Schwinger models with opposite chiralities.

The soldering of the two pieces in (\ref{chiral_ATT}) to reproduce the usual gauge
invariant theory in four dimensions will now be demonstrated.
Consider the soldering transformations

\be
\varphi_a^{(\pm)}= \eta_a;~~\delta \,\mbox{}^*A_a^\mu=0
\label{transff}
\ee
and compute the Noether charges,

\begin{equation}
J_a^{(\pm)} = 2\left(\pm\,\omega\,\varphi_b^{(\pm)}\epsilon_{ba} - \omega^2 \, \varphi_a^{(\pm)}
- g {\cal A}_a^{\pm} \right) 
\label{noether}
\end{equation}
from the variations

\begin{equation}
\delta {\cal S}_{\pm}=J_a^{(\pm)}\eta_a 
\label{variat}
\end{equation}
The effective soldered action in momentum space that is invariant under (\ref{transff}), will be,

\begin{equation}
 {\cal S}_{eff}[\phi_a]={\cal S}_{+}^{(0)}[\varphi^{(+)}_a] + {\cal S}_{-}^{(0)}[\varphi^{(-)}_a]+
 \int d^4k \frac 1{8\omega^2}\left\{ J_a^{(-)}[\varphi^{(+)}_a]+J_a^{(+)}[\varphi^{(-)}_a]\right\}^2
\label{effective}
\end{equation}
where $\phi_a =\varphi_a^{(+)}- \varphi_a^{(-)}$.  Next we perform the inverse Fourier
transform 
to obtain the coordinate space efecctive action as,

\begin{equation}
{\cal S}_{eff}[\phi] = \int d^4x \, \left\{ \frac 12 \left(\partial_\mu \phi\right)^2 -
g\, \partial_\mu \phi \, \epsilon^{\mu\nu\lambda\rho}  A_{\nu\lambda\rho}\right\} +
(a -1) {\cal S}_{reg}.
\end{equation}
Exactly as happened in the two dimensional case, the effective action will be gauge invariant provided,

\begin{equation}
a=1
\label{parameter}\end{equation}
By including the appropriate Maxwell term to impart dynamics to the gauge field,
this is the precise form of the four dimensional Schwinger model as given
in \cite{ATT}.
The analogy with the two dimensional analysis is now complete. 

Just as the
chiral Schwinger models in two dimensions were soldered to yield the
usual Schwinger model, the same has been feasible in four dimensions.
The generation of the massive mode also follows along the same lines.
The observation that not only the different models but the connection 
between them is very similar in two and four dimensions perhaps serves to
demystify the special role that is usually assigned to the former. Indeed our
analysis is completely general applicable to any 2n-dimensions. In other words the
Schwinger model obtained by a straightforward lift can always be thought as an
interference effect among the chiral components analogous to the forms given here.
A natural outcome of the analysis has been the obtention of lagrangians representing
chiral bosons in any even dimensions. Regarding further prospects it would be nice
to actually show the connection between the effective theories proposed here and
chiral $QCD$ in the large N limit, which is expected from plausibility arguments
that relate the bosonised schwinger model in four dimensions
\cite{ATT} with $QCD$ in the large N limit.

\end{document}